\documentclass[aps, preprint, superscriptaddress]{revtex4-2}
\usepackage{graphicx}
\usepackage{makecell, multirow}
\usepackage{color, bm, soul, xcolor}
\usepackage{amsmath, amssymb, amsfonts}
\usepackage{gensymb}
\usepackage{array}
\usepackage{dcolumn}
\usepackage{geometry}
\usepackage{braket}
\usepackage{diagbox}
\usepackage[colorlinks=true, linkcolor=blue, citecolor=blue, urlcolor=blue]{hyperref}
\begin{document}

\title{Interlayer Registry Controls Polar Topology in Twisted Bilayer $h$-BN}

\author{Di Fan}
\affiliation{Department of Physics, School of Science, Westlake University, Hangzhou 310030, China}

\author{Changming Ke}
\email{kechangming@westlake.edu.cn}
\affiliation{Department of Physics, School of Science, Westlake University, Hangzhou 310030, China}
\affiliation{Institute of Natural Sciences, Westlake Institute for Advanced Study, Hangzhou 310024, China}

\author{Shi Liu}
\email{liushi@westlake.edu.cn}
\affiliation{Department of Physics, School of Science, Westlake University, Hangzhou 310030, China}
\affiliation{Institute of Natural Sciences, Westlake Institute for Advanced Study, Hangzhou 310024, China}

\begin{abstract}
Twisting van der Waals bilayers creates a moir\'e stacking field with periodically varying local interlayer registry, enabling spatially patterned polarization. However, a systematic understanding of how this registry-controlled polar topology evolves with twist angle, temperature, and mechanical strain is still lacking. Here, using first-principles-trained deep-potential molecular dynamics simulations, we show that spatially varying interlayer registry governs the polarization texture and can be directly tuned by twist and relative layer deformation. As the twist angle decreases, the moir\'e polarization texture evolves from compact, nearly circular vortex-like patterns to extended triangular AB/BA domains. This crossover arises from enhanced lattice relaxation at small twist angles, which expands low-energy, oppositely polarized AB/BA regions and confines in-plane polarization mainly to domain boundaries. Despite this reconstruction, the local in-plane winding remains tied to interlayer registry and persists up to $1000~\mathrm{K}$, indicating robust topology against thermal fluctuations. We further show that the mechanical response is dictated by whether deformation changes the relative registry: homogeneous deformation of both layers produces only weak distortions, whereas layer-selective deformation strongly reconstructs the moir\'e polarization texture. A mismatch-matrix registry-mapping model is developed to capture this response, explaining deformation-induced changes in vortex radii at larger twist angles and triangular-domain geometry at smaller twist angles. These results identify interlayer registry as the central control parameter for forming and tuning moir\'e polar topology in twisted bilayer $h$-BN.
\end{abstract}

\maketitle

\clearpage

\section*{Introduction}

Introducing a relative twist between van der Waals (vdW) layers creates moir\'e superlattices characterized by long-wavelength periodicity and spatially varying interlayer registry \cite{Geim13p419,Bistritzer11p12233,Andrei20p1265}. These moir\'e superlattices have enabled correlated electronic phases, superconductivity, and spatially confined excitonic states in twisted graphene and transition-metal dichalcogenide heterostructures \cite{Cao18p80,Cao18p43,Seyler19p66,Tran19p71}. Beyond electronic and optical phenomena, the local interlayer registry in vdW bilayers can break inversion symmetry, giving rise to spontaneous electric polarization and moir\'e ferroelectricity in atomically thin crystals \cite{Zheng20p71,Wang22p367,Weston22p390}. Because the polarization is determined by the local registry, polar states can be controlled structurally without changing chemical composition \cite{Bennett22p7}. In twisted bilayers, the interlayer registry varies periodically across the moir\'e pattern, naturally producing spatially textured polarization fields. This registry-driven mechanism provides a route to real-space polar topology distinct from that in ferroelectric nanostructures and oxide heterostructures, where polar vortices, skyrmions, and meron-like textures typically arise from competing electrostatic, elastic, and geometric constraints \cite{Junquera23p025001,Hong26p1022}. This distinction raises the question of whether the polarization textures originating from the interlayer-registry field in twisted vdW bilayers exhibit unique features.

Twisted bilayer $h$-BN provides a particularly clean platform for examining how interlayer registry organizes polarization textures.     
In parallel-aligned bilayer $h$-BN, AB and BA stackings are connected by lateral sliding and carry opposite out-of-plane polarizations \cite{Li17p6382,Yasuda21p1458,ViznerStern21p1462}. 
Twist converts the uniform sliding coordinate into a spatially varying interlayer-registry field containing AB, BA, and AA-like regions \cite{Ni19p4360,woods21p347,moore21p5741}. 
Experiments on marginally twisted $h$-BN have shown that lattice reconstruction and electromechanical effects like flexoelectricity can substantially reshape moir\'e ferroelectric domains \cite{Kim24p65,Wan24p2410563}.  
Theoretical studies have further predicted meron--antimeron topology in the polar domain network, arising from the in-plane component of the local polarization \cite{Bennett22p235445,Bennett23p1629}. Recent experiments have reported spatially resolved polarization textures with varying in-plane and out-of-plane polarization components in moir\'e ferroelectrics \cite{Pan25p3026,li25p5451}.

These advances establish interlayer registry and lattice relaxation as central structural ingredients of moir\'e polarization textures. Yet how registry-organized topology evolves under external stimuli, such as temperature and mechanical deformation, remains less explored. In particular, many previous studies of strain effects in moir\'e superlattices have considered deformations applied equivalently to both layers, which largely preserve the relative registry. Because the moir\'e length scale is set by a small relative mismatch, even a modest deformation difference between the two layers can substantially alter the registry gradient and reconstruct the moir\'e pattern. We refer to such inequivalent deformations of the two layers as \textit{layer-selective deformation}, which is a main focus of this work.

Here, using first-principles-trained deep-potential molecular dynamics (DPMD) simulations, we investigate how temperature and layer-selective deformation affect the evolution of polarization textures in twisted bilayer $h$-BN. We find that, at larger twist angles, the short moir\'e period produces small AB and BA regions in which most local unit cells develop both out-of-plane polarization and substantial in-plane polarization, with the in-plane component forming nearly circular vortex-like textures. As the twist angle decreases and the moir\'e length grows, lattice relaxation enlarges the AB and BA domains. Their interiors become dominated by out-of-plane polarization with weak in-plane components, while sizable in-plane polarization localizes near domain boundaries, forming triangular vortex-like textures. Importantly, in all investigated twisted bilayers, the in-plane winding number depends weakly on temperature and persists up to $1000~\mathrm{K}$, suggesting that registry-organized topology is more robust than polarization textures in perovskite superlattices based on delicate energetic competition. Layer-selective deformation substantially modifies the moir\'e pattern and stabilizes diverse polar networks, including elongated, compressed, and stripe-like textures similar to experimentally reported strain-distorted moir\'e domains \cite{Guan25p986}. We further develop a geometric registry-mapping model that analytically explains the deformation of moir\'e patterns observed in MD simulations. Our results show that registry-organized polarization textures are robust against thermal fluctuations and common deformation that largely preserve the relative registry, but highly sensitive to layer-selective deformation that directly modifies the registry field. 

\section*{Results}
\subsection*{Twist-angle-dependent polarization textures}

\begin{figure}[t]
\includegraphics[width=0.7\textwidth]{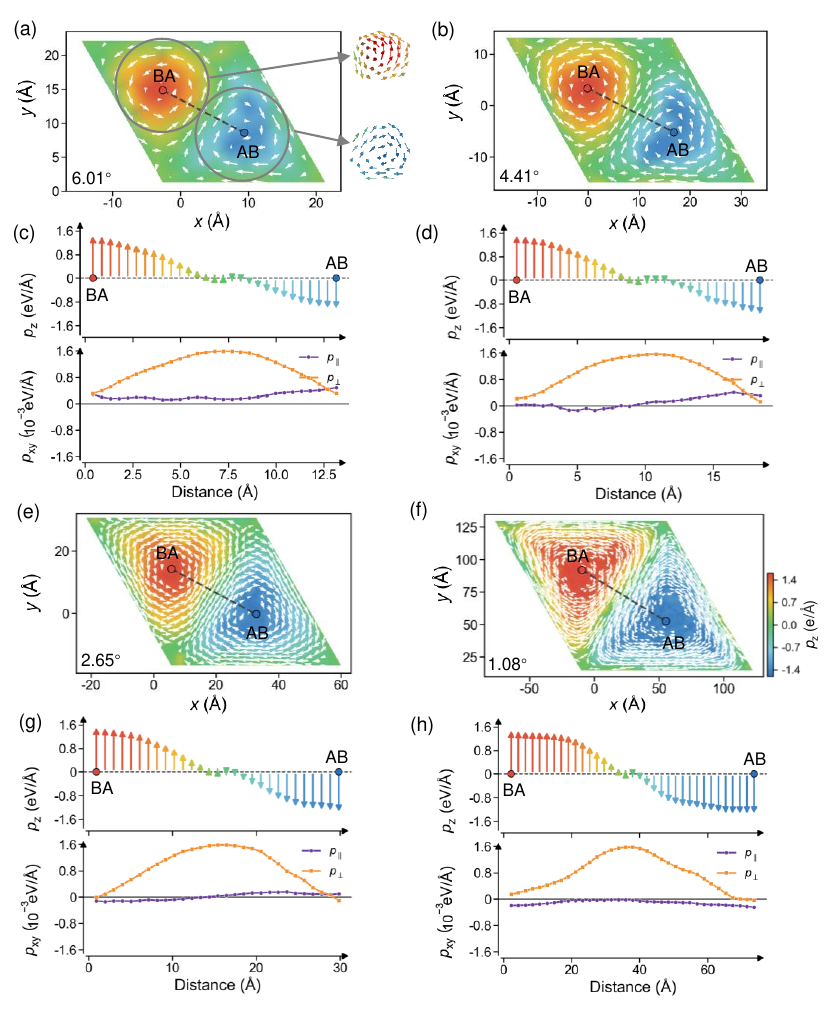}
\caption{\textbf{Twist-angle evolution of polar topology in twisted bilayer $h$-BN.}
(\textbf{a}, \textbf{b}, \textbf{e}, \textbf{f}) Representative polarization textures at twist angles of $6.01^\circ$, $4.41^\circ$, $2.65^\circ$, and $1.08^\circ$, respectively. The background colors encode $P_z$, while the white arrows represent $\mathbf{P}_{xy}$.
(\textbf{c}, \textbf{d}, \textbf{g}, \textbf{h}) Corresponding polarization profiles along the AB--BA pathway at the same four twist angles, respectively. The profiles show the reversal of $P_z$ between opposite polar domains; the lower traces give the components of $\mathbf{P}_{xy}$ parallel and perpendicular to the AB--BA path. The predominance of the path-perpendicular component in the switching region indicates a Bloch-like rotation of the polarization at the AB--BA interfaces.}
\label{fig:polar_texture}
\end{figure}

We first examine how the polarization texture evolves with twist angle. Figure~\ref{fig:polar_texture}(a,b,e,f) shows that AB and BA regions with opposite out-of-plane polarizations persist across the representative twist-angle series. At larger twist angles, such as $6.01^\circ$ and $4.41^\circ$ [Figs.~\ref{fig:polar_texture}(a,b)], the moir\'e cell contains relatively small AB and BA regions. Most local unit cells in these regions carry both an out-of-plane polarization $P_z$ and a sizable in-plane component $\mathbf{P}_{xy}=(P_x,P_y)$. The in-plane polarization forms nearly circular vortex-like textures around the oppositely polarized AB and BA regions. In the AB region, where $P_z<0$, the in-plane polarization winds clockwise, whereas in the BA region, where $P_z>0$, it winds counterclockwise. This coupling between the sign of $P_z$ and the handedness of the in-plane vortex remains identifiable as the twist angle is reduced.

With decreasing twist angle, lattice relaxation becomes increasingly important as the system lowers its energy by expanding nearly commensurate, low-energy AB/BA domains. As a result, the centers of the AB and BA domains become dominated by out-of-plane polarization and exhibit much smaller in-plane components, while the sizable in-plane polarization is pushed toward the AB/BA boundaries. As shown in Fig.~\ref{fig:polar_texture}(f), at $1.08^\circ$ the in-plane polarization is primarily concentrated along the domain boundaries, forming a triangular vortex-like texture whose rotation direction is locked to the out-of-plane polarization. The AB and BA domains retain opposite winding directions, consistent with the same $P_z$--handedness coupling observed at larger twist angles.

\begin{figure}[t]
\includegraphics[width=0.75\textwidth]{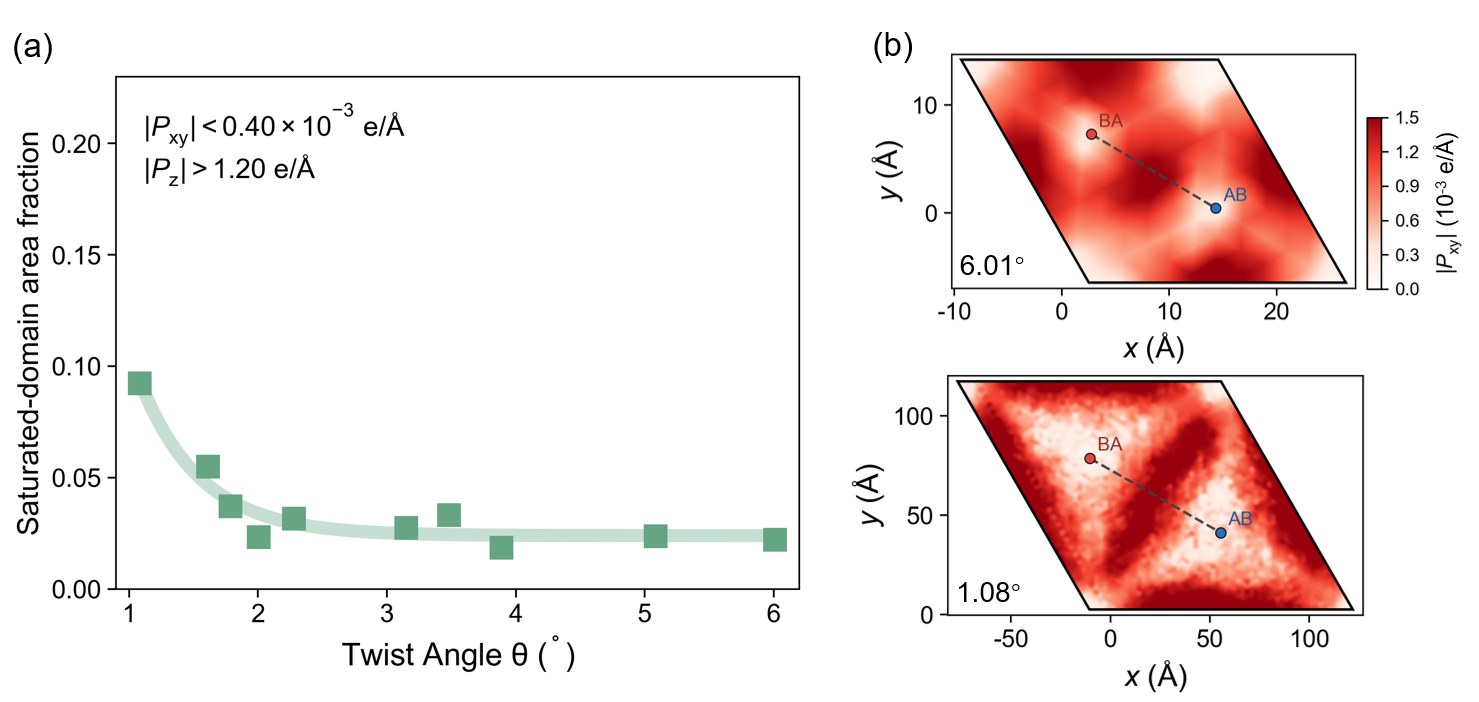}
\caption{\textbf{Twist-angle evolution of the saturated-domain area fraction and in-plane polarization magnitude in twisted bilayer $h$-BN.} (\textbf{a}) Twist-angle dependence of the saturated-domain area fraction, defined as the fraction of the moir\'e unit cell satisfying $|\mathbf{P}_{xy}|<0.40\times10^{-3}$~e/\AA~and $|P_z|>1.20$~e/\AA. (\textbf{b}) Representative spatial distributions of the in-plane polarization magnitude $|\mathbf{P}_{xy}|$ at twist angles of $6.01^\circ$ and $1.08^\circ$, shown using a common color scale.}
\label{fig:domain_fraction}
\end{figure}

The enhanced lattice relaxation at small twist angles is further quantified in Fig.~\ref{fig:domain_fraction}(a), which measures the fraction of the moir\'e unit cell exhibiting weak in-plane polarization and nearly saturated out-of-plane polarization characteristic of ideal commensurate AB or BA stackings. Specifically, the saturated-domain area fraction is defined as the portion of the unit cell satisfying $|\mathbf{P}_{xy}|<0.40\times10^{-3}$~e/\AA~and $|P_z|>1.20$~e/\AA. This fraction increases overall as the twist angle decreases, with a particularly pronounced increase for $\theta<2^\circ$. Figure~\ref{fig:domain_fraction}(b) compares representative maps of $|\mathbf{P}_{xy}|$ at $6.01^\circ$ and $1.08^\circ$ using the same color scale. At $6.01^\circ$, the low-$|\mathbf{P}_{xy}|$ regions are confined mainly to the centers of the AB, BA, and AA regions. In contrast, at $1.08^\circ$, the low-$|\mathbf{P}_{xy}|$ regions occupy a much larger fraction of the moir\'e cell and develop a triangular shape, reflecting the expansion of relaxed AB/BA domains.

To further clarify the polarization texture, we examine the polarization profiles along the AB--BA pathway in Fig.~\ref{fig:polar_texture}(c,d,g,h). For all twist angles, $P_z$ reverses between the oppositely polarized BA and AB regions. The in-plane polarization in the switching region is dominated by the path-perpendicular component, $P_{\perp}$, whereas the path-parallel component, $P_{\parallel}$, remains comparatively small. During the reversal of $P_z$, the in-plane polarization therefore develops mainly along the domain-wall direction. This behavior is characteristic of a Bloch-type domain wall, where the polarization rotates between opposite out-of-plane orientations through an in-plane component parallel to the wall.

Overall, the twist-angle evolution reflects a competition between the moir\'e-imposed interlayer-registry field and lattice relaxation. Large-angle structures exhibit smaller AB/BA domains, where incomplete lattice relaxation likely gives rise to sizable in-plane winding components. In contrast, small-angle structures develop extended AB and BA domains whose interiors are dominated by out-of-plane polarization, with the in-plane components pushed toward the boundaries. For all investigated twist angles, the domain walls separating the BA and AB regions show clear Bloch-type features, with in-plane components parallel to the walls. Thus, the circular and triangular vortex-like textures can be viewed as two morphological limits of the same registry-controlled polar topology.

\subsection*{Thermal stability of the polar winding}

We next examine the persistence of the in-plane polarization winding under thermal fluctuations. For a closed contour \(C\) enclosing an individual polar core, defined as the center of an AB or BA domain, the winding of the in-plane polarization is evaluated as
\begin{equation}
Q_{xy}=\frac{1}{2\pi}\oint_C d\phi,
\qquad
\phi=\arg(P_x+iP_y),
\end{equation}
where \(\phi\) is the azimuthal angle of the in-plane polarization. Thus, $Q_{xy}$ measures the net number of $2\pi$ rotations of the in-plane polarization along the contour. For an ideal vortex-like texture, $Q_{xy}=\pm1$ corresponds to one complete circulation of $\mathbf{P}_{xy}$ around the core, with the sign determined by the winding direction.

\begin{figure}[t]
\includegraphics[width=0.6\textwidth]{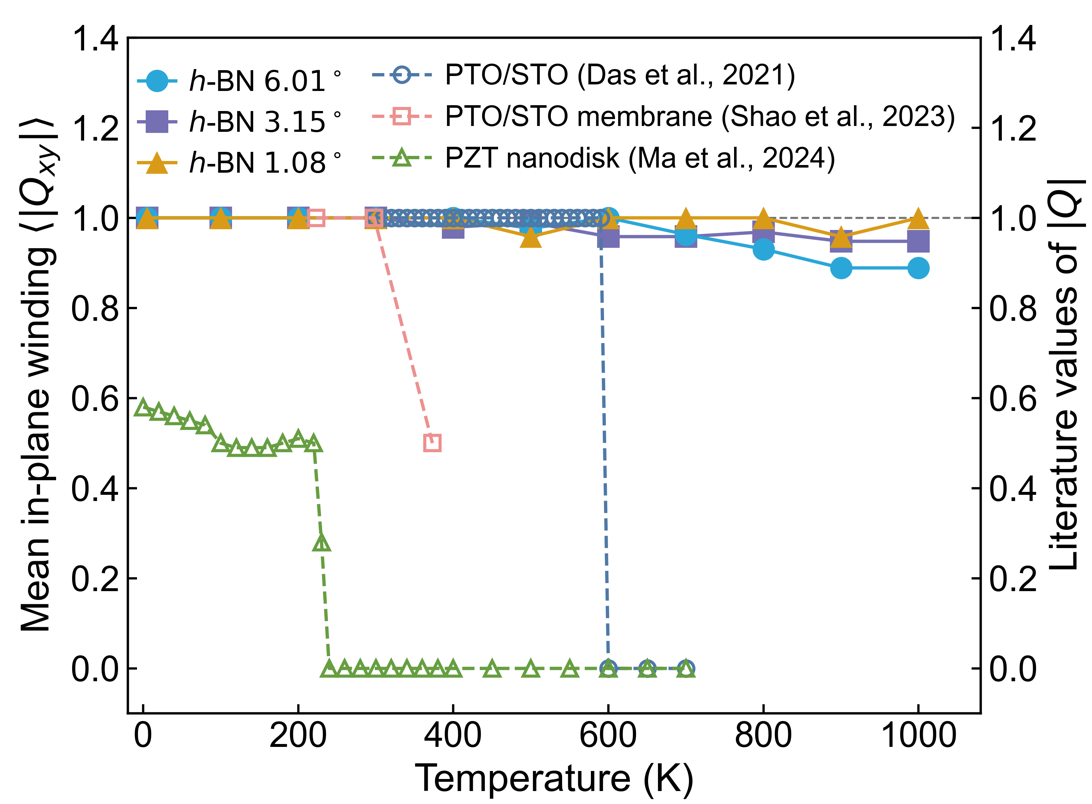}
\caption{\textbf{Thermal persistence of the in-plane polarization winding in twisted bilayer $h$-BN.} Temperature dependence of the mean absolute in-plane polarization winding, $\langle |Q_{xy}| \rangle$, for twisted bilayer $h$-BN at twist angles of $6.01^\circ$, $3.15^\circ$, and $1.08^\circ$, together with representative temperature-dependent polar topological observables reported for PbTiO$_3$/SrTiO$_3$ superlattices, PbTiO$_3$/SrTiO$_3$ membranes, and Pb(Zr,Ti)O$_3$ nanodisks \cite{Das21p194,Shao2023p1355,Ma24p224105}. For $h$-BN, the plotted quantity is the mean absolute winding number, $\langle |Q_{xy}| \rangle$, evaluated from the in-plane polarization winding around individual polar cores and averaged over configurations sampled during the final $100$~ps of each $300$~ps trajectory. The Das et al.~\cite{Das21p194} data show the experimental $|N_Q|$, the Shao et al.~\cite{Shao2023p1355} open squares denote reported skyrmion and meron states at selected temperatures, and the Ma et al.~\cite{Ma24p224105} curve is digitized from the reported $Q(T)$ data. The oxide data are included as representative temperature-dependent trends for qualitative comparison.}
\label{fig:thermal_stability}
\end{figure}

Figure~\ref{fig:thermal_stability} shows the mean absolute winding, $\langle |Q_{xy}| \rangle$, for twisted bilayer $h$-BN at twist angles of $6.01^\circ$, $3.15^\circ$, and $1.08^\circ$. Over the investigated temperature range, $\langle |Q_{xy}| \rangle$ remains close to unity, indicating that the local in-plane polarization winding persists despite thermal fluctuations. At $1000~\mathrm{K}$, the corresponding mean values are approximately $0.88$, $0.94$, and $1.00$, respectively. The small variations at elevated temperatures, most evident for the larger-angle structures, arise from thermal distortions of the instantaneous polarization textures. Within the simulated trajectories and analysis window, no sustained loss of the winding pattern is detected. 

For qualitative context, Fig.~\ref{fig:thermal_stability} also shows representative temperature-dependent topological observables reported for oxide polarization textures~\cite{Das21p194,Shao2023p1355,Ma24p224105}. These studies show that polar skyrmion, meron, and vortex textures in oxides can undergo temperature-driven structural transformations or disappear at elevated temperatures.
We suggest that the robustness of polar topology in  twisted bilayer $h$-BN originates from the registry-controlled nature. The moir\'e geometry fixes the spatial sequence of AB, BA, and intermediate stackings, which determines both the sign of $P_z$ and the winding pattern of $\mathbf{P}_{xy}$. Thermal fluctuations can locally distort the polarization field, but they do not provide a directional driving force to change the registry sequence.
In contrast, the stability of many oxide topological polar structures relies on a delicate balance among polarization, electrostatic energy, depolarization fields, and polarization--strain coupling. These competing interactions can be strongly modified by temperature, making oxide polarization textures more susceptible to thermally driven transformations.

\subsection*{Registry mapping under twist and layer-selective deformation}

Mechanical deformation provides an additional route for tuning moir\'e polarization textures. As a reference case, we first considered common homogeneous deformation, in which the same deformation matrix is applied to both $h$-BN layers. This operation largely preserves the relative interlayer registry and produces only minor changes in the polarization texture. In contrast, layer-selective deformation, in which a deformation is applied to one layer while the other layer is kept fixed, directly changes the relative registry field and therefore reshapes the moir\'e polarization texture much more effectively. We therefore focus on layer-selective deformation below.

Here, we develop a geometric registry-mapping model to relate the polar-texture reconstruction to changes in the interlayer registry field. To this end, we first describe how the local registry is determined in a twisted bilayer, since this registry directly controls the polar topology. We start from an untwisted, perfectly aligned $h$-BN bilayer, where the two layers share the same lattice orientation. The twisted bilayer is constructed by rotating the two layers by $+\theta/2$ and $-\theta/2$, giving a relative twist angle $\theta$. For describing the local stacking, this is equivalent to taking the lower-layer lattice as the reference, relative to which the upper layer is rotated
by $+\theta$. We denote the corresponding rotation matrix by $\mathbf R_\theta$.

At a given physical position $\mathbf r$ in the twisted bilayer, the local registry is determined by comparing the lattice phases associated with the two layers at this same position $\mathbf r$. Here, the ``lattice phase'' $\mathbf s$ labels the corresponding position in the original untwisted $h$-BN lattice.  It can be viewed as a fractional coordinate within the primitive unit cell, with phases that differ by an integer Bravais lattice identified as equivalent. For example, B, N, and hollow sites have different fractional coordinates and therefore different lattice phases. Twisting changes the physical positions of lattice sites, but not their lattice phases.

We introduce the primitive lattice matrix $\mathbf A=(\mathbf a_1,\mathbf a_2)$ of the $h$-BN unit cell, where $\mathbf a_1$ and $\mathbf a_2$ are the primitive lattice vectors. The inverse matrix $\mathbf A^{-1}$ maps a physical position
to the corresponding lattice phase. For a position $\mathbf r_0$ in the reference frame fixed to the lower layer, the lower-layer phase is
$\mathbf s_{\mathrm l}(\mathbf r_0)=\mathbf A^{-1}\mathbf r_0$, where the result is implicitly taken modulo integer vectors. That is, two physical positions that differ by a Bravais lattice vector have the same lattice phase, as expected from the periodicity of the $h$-BN lattice.
To find the upper-layer phase at the same physical position $\mathbf r_0$, we first map $\mathbf r_0$ back to the untwisted upper-layer frame, which has the same orientation as the lower layer. This is done by applying the inverse rotation $\mathbf R_{-\theta}$ and then converting to lattice phase:
\begin{equation}
\mathbf s_{\mathrm u}(\mathbf r_0)
=
\mathbf A^{-1}\mathbf R_{-\theta}\mathbf r_0 .
\end{equation}
The relative phase is defined as the difference between the two lattice phases,
\begin{equation}
\boldsymbol{\eta}_0(\mathbf r_0)
=
\mathbf s_{\mathrm l}(\mathbf r_0)-\mathbf s_{\mathrm u}(\mathbf r_0),
\end{equation}
with $\boldsymbol{\eta}_0$ taken modulo integer vectors. Multiplying by $\mathbf A$ gives the corresponding real-space stacking vector,
\begin{equation}
\boldsymbol{\delta}_0(\mathbf r_0)
=
\mathbf A\boldsymbol{\eta}_0(\mathbf r_0)
=
\left(\mathbf I-\mathbf R_{-\theta}\right)\mathbf r_0 .
\end{equation}
This vector characterizes the local registry at position $\mathbf r_0$ in a bilayer with twist angle $\theta$.

We next apply a layer-selective deformation $\mathbf F$ to the upper layer after the twist, while keeping the lower layer fixed. Here, $\mathbf F$ is a $2\times2$ deformation matrix defined in the fixed Cartesian frame of the lower layer, incorporating the uniaxial strain $\epsilon$ and shear strain $\gamma$.
The main idea is the same as above: at a given physical position after deformation, we determine the upper-layer lattice phase by mapping that position back to the undeformed, untwisted upper-layer lattice. The local registry is then obtained by comparing this phase with the lower-layer phase at the same position. 
At a physical position $\mathbf r_s$ in the strained twisted bilayer, the upper-layer phase is obtained by undoing the deformation, undoing the twist, and then converting to lattice phase:
\begin{equation}
\mathbf s_{\mathrm u}(\mathbf r_s)
=
\mathbf A^{-1}\mathbf R_{-\theta}\mathbf F^{-1}\mathbf r_s .
\end{equation}
Since the lower layer is fixed, its phase remains
$\mathbf s_{\mathrm l}(\mathbf r_s)=\mathbf A^{-1}\mathbf r_s$. Therefore, the relative registry phase is
\[
\boldsymbol{\eta}_s(\mathbf r_s)
=
\mathbf s_{\mathrm l}(\mathbf r_s)-\mathbf s_{\mathrm u}(\mathbf r_s)
=
\mathbf A^{-1}
\left(
\mathbf I-\mathbf R_{-\theta}\mathbf F^{-1}
\right)
\mathbf r_s ,
\]
taken modulo integer vectors. Equivalently, the corresponding real-space stacking vector is
\begin{equation}
\boldsymbol{\delta}_s(\mathbf r_s)
=
\mathbf A\boldsymbol{\eta}_s(\mathbf r_s)
=
\left(
\mathbf I-\mathbf R_{-\theta}\mathbf F^{-1}
\right)
\mathbf r_s .
\end{equation}

This expression allows us to track how a given local registry moves under the layer-selective deformation. Let $\mathbf r_0$ be the position of a given registry in the unstrained twisted bilayer, and let $\mathbf r_s$ be the position of the same registry after deformation. Since the registry is the same, the stacking vectors must be equal modulo primitive lattice vectors:
\[
\boldsymbol{\delta}_0(\mathbf r_0)=\boldsymbol{\delta}_s(\mathbf r_s),
\]
which further gives
\begin{equation}
\left(\mathbf I-\mathbf R_{-\theta}\right)\mathbf r_0
=
\left(
\mathbf I-\mathbf R_{-\theta}\mathbf F^{-1}
\right)
\mathbf r_s .
\end{equation}
Defining 
\[
\mathbf M_0=\mathbf I-\mathbf R_{-\theta},
\qquad
\mathbf M_s=\mathbf I-\mathbf R_{-\theta}\mathbf F^{-1},
\]
the registry matching condition becomes $\mathbf M_0\mathbf r_0=\mathbf M_s\mathbf r_s$. If $\mathbf M_s$ is invertible, then the mapping from the unstrained to the strained moir\'e pattern is 
\begin{equation}
\mathbf r_s
=\mathbf T\mathbf r_0
\end{equation}
with 
\begin{equation}
\mathbf T
=
\mathbf M_s^{-1}\mathbf M_0
=
\left(
\mathbf I-\mathbf R_{-\theta}\mathbf F^{-1}
\right)^{-1}
\left(
\mathbf I-\mathbf R_{-\theta}
\right).
\label{eq:registry_mapping}
\end{equation}
Thus, $\mathbf T$ maps the position of a given local registry in the unstrained twisted bilayer to the position of the same local registry after the upper-layer deformation.

This mapping provides a direct geometric description of how the polarization texture is deformed. For example, consider the in-plane vortex texture in Fig.~\ref{fig:polar_texture}a. In the unstrained structure, take a circle of radius $R_0$ centered on the vortex,
\[
\mathbf r_0^{\mathrm{T}}\mathbf r_0 = R_0^2 ,
\]
where $\mathbf r_0^{\mathrm{T}}$ denotes the transpose of the position vector $\mathbf r_0$.
After applying $\mathbf F$ to the upper layer, points with the same local registry are mapped to $\mathbf r_s=\mathbf T\mathbf r_0$. Substituting $\mathbf r_0=\mathbf T^{-1}\mathbf r_s$ gives
\begin{equation}
\left(\mathbf T^{-1}\mathbf r_s\right)^{\mathrm T}
\left(\mathbf T^{-1}\mathbf r_s\right)
=
R_0^2 ,
\end{equation}
which describes an ellipse in the strained moir\'e pattern. The deformation of the vortex texture can therefore be quantified from the directional radii of this ellipse. Denoting the effective radius along the direction $\alpha=x,y$ by $R_\alpha$, and substituting $\mathbf r_s=(R_x,0)$ or $\mathbf r_s=(0,R_y)$ into the ellipse equation, we obtain
\begin{equation}
\frac{R_\alpha}{R_0}
=
\frac{1}{
\sqrt{
\left[
\left(\mathbf T\mathbf T^{\mathrm T}\right)^{-1}
\right]_{\alpha\alpha}}},
\qquad
\alpha=x,y .
\label{eq:geometric_radius}
\end{equation}

We can evaluate Eq.~\eqref{eq:registry_mapping} and Eq.~\eqref{eq:geometric_radius} using a small-parameter approximation for $\theta\ll1$, $|\gamma|\ll1$, and $|\epsilon|\ll1$, with all angles expressed in radians. For example, for the $S_x$ mode (see Methods), the deformation matrix is
\begin{equation}
\mathbf F_{S_x}=
\begin{pmatrix}
1 & \gamma\\
0 & 1
\end{pmatrix}.
\end{equation}
This gives
\[
\frac{R_{x}}{R_0}\approx 1,
\qquad
\frac{R_{y}}{R_0}
\approx \frac{\theta}{\theta-\gamma}.
\]

\subsection*{Layer-selective deformation of polar vortices at large twist angles}

\begin{figure}[t]
\includegraphics[width=1\textwidth]{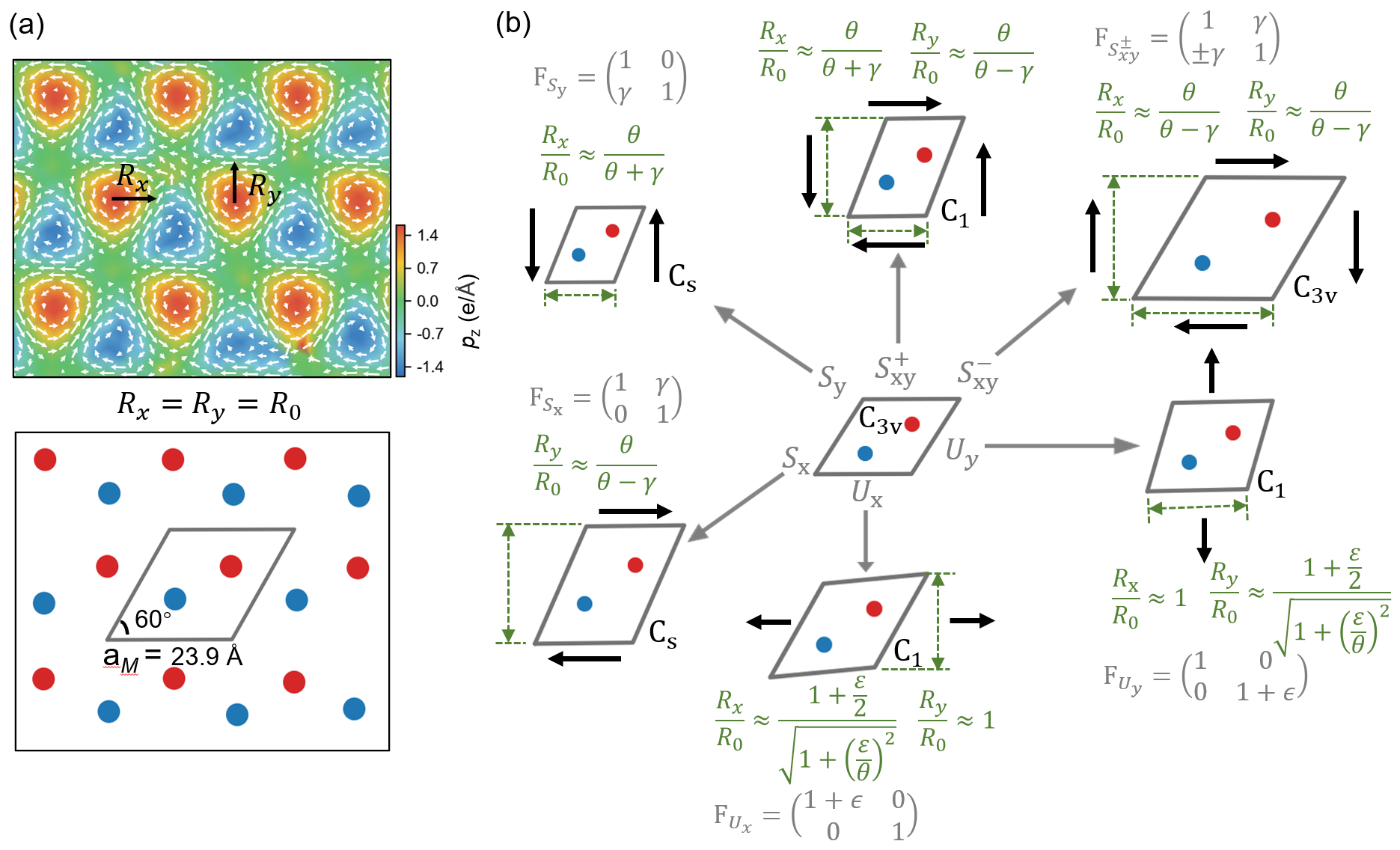}
\caption{\textbf{Layer-selective deformation reconstructs the moir\'e vortex lattice in twisted bilayer $h$-BN.}
(\textbf{a}) Polarization texture of the unstrained twisted bilayer $h$-BN with  $\theta=6.01^{\circ}$ and the corresponding reduced vortex-core lattice. The color denotes $P_z$ using the same scale as in Fig.~\ref{fig:polar_texture}(f), the arrows denote $\mathbf{P}_{xy}$, and the red and blue points represent upward- and downward-polarized cores, respectively. 
(\textbf{b}) Qualitative moir\'e-cell responses to the layer-selective deformation modes $S_x$, $S_y$, $S_{xy}^{+}$, $S_{xy}^{-}$, $U_x$, and $U_y$. The black arrows indicate the imposed upper-layer deformation, and the gray arrows indicate the resulting cell reconstruction. The schematics correspond to positive deformation amplitudes. The green expressions are simplified leading-order forms derived in the small-angle and small-deformation limit, where \(\theta\) denotes the interlayer twist angle, \(\gamma\) the signed shear amplitude, and \(\epsilon\) the engineering strain. Angular quantities are expressed in radians; the expressions are shown only to illustrate the underlying geometric trends. The symmetry labels indicate the approximate point symmetry of the reconstructed vortex lattice.
}
\label{fig:strain_lattice}
\end{figure}

We first consider an unstrained bilayer with a relatively large twist angle of $\theta=6.01^{\circ}$ to demonstrate the effect of layer-selective deformation. At this twist angle, the in-plane polarization forms a vortex array, and each vortex is characterized by two geometric parameters, $R_x$ and $R_y$, as shown in Fig.~\ref{fig:strain_lattice}a. Here, ``relatively large'' is used only in comparison with the small-twist-angle case discussed in Fig.~\ref{fig:polar_texture}, where larger AB/BA domains appear. Nevertheless, the twist angle remains small in radians, $\theta=0.105$, so the deformation can still be described by the registry-mapping model.
Representing each vortex by its polar core gives a nearly hexagonal point lattice, with moir\'e lattice constant $a_M\approx23.9~\text{\AA}$ and angle $\phi_M=60^{\circ}$ between the two moir\'e lattice vectors. This reduced representation removes the internal polarization profile while retaining the vortex positions and moir\'e lattice symmetry.

Figure~\ref{fig:strain_lattice}b illustrates the six loading modes, their corresponding deformation matrices, and the radius ratios $R_\alpha/R_0$ derived from the registry-mapping model discussed in the previous section. The uniaxial modes $U_x$ and $U_y$ mainly modify the cell angle, reduce the point-group symmetry to $C_1$, and produce only weak changes in the vortex size, with $R_x\approx R_y\approx R_0$ when $\epsilon$ and $\epsilon/\theta$ are both small. The single-axis shear modes $S_x$ and $S_y$, by contrast, selectively change one directional radius $R_\alpha$, while leaving the other approximately equal to $R_0$.
The symmetric coupled shear mode $S_{xy}^{+}$ enhances the anisotropy between the two directional vortex radii, giving $R_y>R_x$. In contrast, the opposite-sign mode $S_{xy}^{-}$ is rotation-like to first order and primarily changes the effective relative twist angle, thereby changing the vortex radius while preserving $R_x=R_y$.

\begin{figure}[t]
\includegraphics[width=1\textwidth]{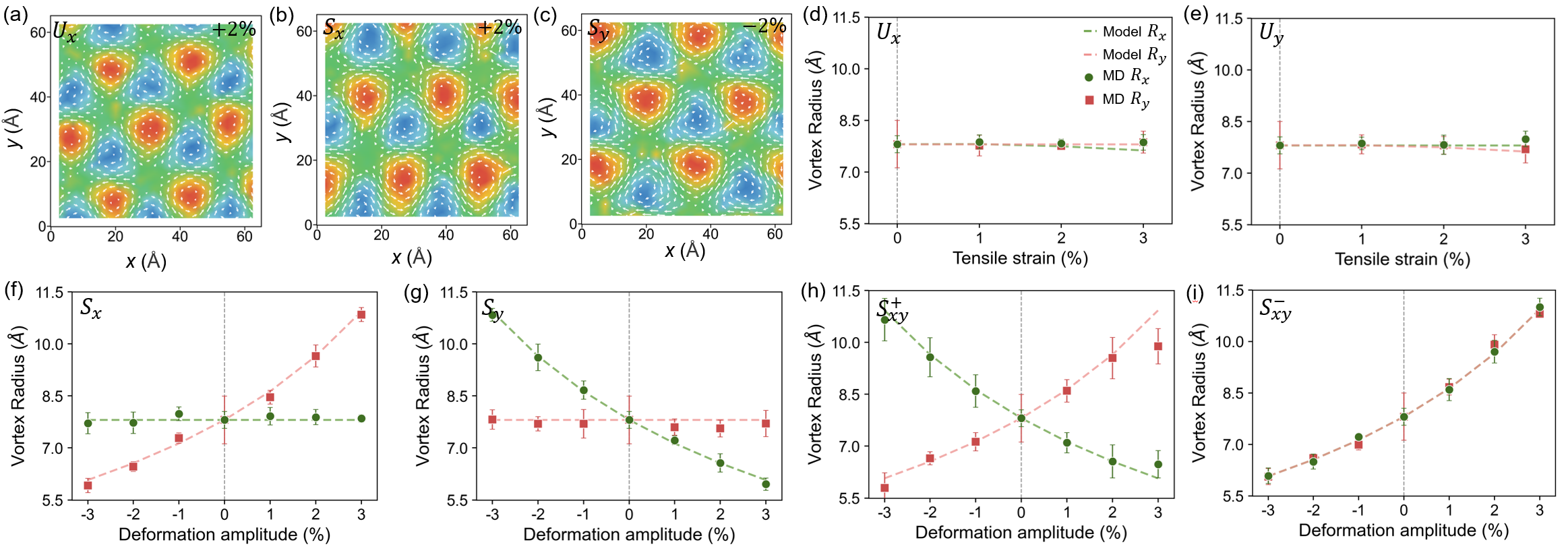}
\caption{\textbf{Polarization textures and directional vortex radii under layer-selective deformation.}
(\textbf{a--c}) Representative polarization distributions for the deformation states
$U_x=+2\%$, $S_x=+2\%$, and $S_y=-2\%$, respectively. Arrows indicate the in-plane polarization $\mathbf{P}_{xy}$, while the color scale represents the out-of-plane polarization $P_z$.
(\textbf{d--i}) Directional vortex radii for the deformation modes $U_x$, $U_y$, $S_x$, $S_y$, $S_{xy}^{+}$, and $S_{xy}^{-}$, respectively. Dashed green and red curves denote the predictions of the full mismatch-matrix registry-mapping model, evaluated from Eqs.~\eqref{eq:registry_mapping}--\eqref{eq:geometric_radius} without using the leading-order approximations shown in Fig.~\ref{fig:strain_lattice}b. Green circles and red squares show the corresponding MD values for $R_x$ and $R_y$, respectively, averaged over six vortex cores; error bars
denote the sample standard deviations across these cores. The radii are determined from the first zero crossing of sector-averaged out-of-plane polarization profiles $P_z$ along the fixed Cartesian $x$ and $y$ axes. The unstrained reference radius is $R_0=7.81$~\AA.}
\label{fig:strain_radius}
\end{figure}

Figure~\ref{fig:strain_radius}(a--c) shows representative MD-simulated polarization textures for $U_x=+2\%$, $S_x=+2\%$, and $S_y=-2\%$, respectively, providing a visual picture of the deformation-induced ellipticity. These patterns show that layer-selective deformation changes the shape of the polar vortex texture. The nearly circular vortex in the unstrained structure becomes elongated under shear deformations such as $S_x$ and $S_y$, while the change is weaker under the uniaxial deformation $U_x$.
We then quantify this effect using the directional vortex radii $R_x$ and $R_y$ extracted from the MD simulations. The unstrained reference radius, determined from the first zero crossing of the out-of-plane polarization profile $P_z$, is $R_0=7.81~\text{\AA}$. The radii $R_x$ and $R_y$ are measured along the fixed Cartesian $x$ and $y$ directions, respectively, and the reported values are averaged over the six vortex cores in each MD simulation supercell.
The uniaxial modes $U_x$ and $U_y$ produce much smaller changes in $R_x$ and $R_y$ over the same strain range [Fig.~\ref{fig:strain_radius}(d,e)]. This weak response is again consistent with the registry-mapping model, which predicts only small changes in the vortex radius when $\epsilon$ and $\epsilon/\theta$ are both small.

The single-axis shear modes show a strong direction-selective response. For $S_x$, $R_x$ remains close to $R_0$, while $R_y$ changes strongly, increasing to $10.85~\text{\AA}$ at $+3\%$ and decreasing to $5.92~\text{\AA}$ at $-3\%$ [Fig.~\ref{fig:strain_radius}(f)]. For $S_y$, the behavior is reversed: $R_y$ remains nearly unchanged, while $R_x$ increases to $10.84~\text{\AA}$ at $-3\%$ and decreases to $5.95~\text{\AA}$ at $+3\%$ [Fig.~\ref{fig:strain_radius}(g)]. These MD results agree well with the registry-mapping model, which predicts that single-axis shear mainly changes one directional radius while leaving the other nearly unchanged.

The $S_{xy}^{+}$ mode produces a clear anisotropic response, with $R_x=6.48~\text{\AA}$ and $R_y=9.89~\text{\AA}$ at $+3\%$ [Fig.~\ref{fig:strain_radius}(h)], consistent with the model prediction that $S_{xy}^{+}$ enhances the difference between the two directional radii.
In contrast, for $S_{xy}^{-}$, the two radii change together, reaching approximately $11~\text{\AA}$ at $+3\%$ and $6~\text{\AA}$ at $-3\%$ [Fig.~\ref{fig:strain_radius}(i)]. This again agrees with the registry-mapping prediction that $S_{xy}^{-}$ acts mainly as a rotation-like deformation and produces an approximately isotropic rescaling of the vortex size.
Overall, the MD simulations agree well with the registry-mapping model. Both show that layer-selective deformation reshapes the moir\'e polarization texture mainly through changes in the interlayer registry. 

\subsection*{Layer-selective deformation of triangular domains at small twist angles}

We next investigate layer-selective deformation in a bilayer with a small twist angle of $\theta=1.08^{\circ}$. This system features triangular AB and BA domains with out-of-plane polarization, while the in-plane polarization forms triangular vortex-like textures along the domain boundaries.
To analyze the domain geometry, we use the AA-point network to define a representative reference triangle. Let $\mathbf a_b$ denote the selected AA--AA base vector, and let $\mathbf a_t$ connect the same AA point to the third vertex of the triangle in the strain-free bilayer. Under layer-selective deformation, these vectors transform as
\begin{equation}
\mathbf a_b'=\mathbf T\mathbf a_b,
\qquad
\mathbf a_t'=\mathbf T\mathbf a_t .
\end{equation}
The base length and the corresponding perpendicular height of the triangular domain are then given by
\begin{equation}
b=|\mathbf a_b'|,
\qquad
h=
\frac{|\mathbf a_b'\times\mathbf a_t'|}
     {|\mathbf a_b'|}.
\end{equation}
Using the registry-mapping model, we can analytically predict $b$ and $h$ for any layer-selective deformation $\mathbf F$. For comparison, we also perform MD simulations to obtain $b$ and $h$ numerically under different deformations, with the corresponding vectors tracked continuously from the strain-free configuration.

\begin{figure}[t]
\includegraphics[width=1\textwidth]{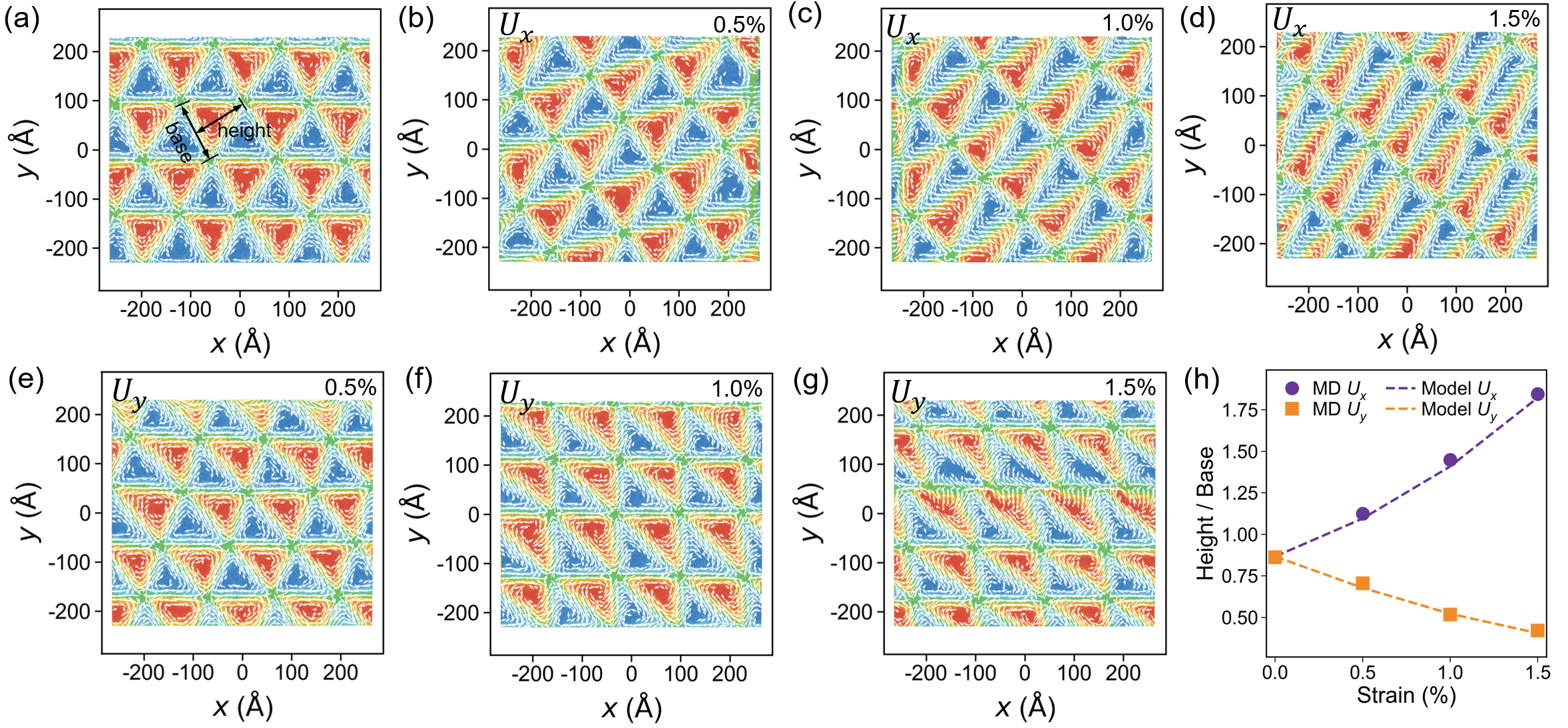}

\caption{\textbf{Uniaxial deformation of triangular polar domains in small-angle twisted bilayer $h$-BN.}
(\textbf{a}) Zero-strain triangular-domain polarization texture in twisted bilayer $h$-BN with $\theta=1.08^{\circ}$.
(\textbf{b--d}) Polarization textures under $U_x$ strains of $0.5\%$, $1.0\%$, and $1.5\%$, respectively.
(\textbf{e--g}) Polarization textures under $U_y$ strains of $0.5\%$, $1.0\%$, and $1.5\%$, respectively.
(\textbf{h}) Triangular-domain aspect ratio $h/b$ as a function of uniaxial strain. Symbols denote MD results, while dashed lines show predictions from the registry-mapping model. The common zero-strain point is shown for both deformation modes.
The color scale for $P_z$ is the same as that used in Fig.~\ref{fig:polar_texture}(f), and the arrows represent the in-plane polarization $\mathbf{P}_{xy}$.}
\label{fig:small_angle_uniaxial}
\end{figure}

Figure~\ref{fig:small_angle_uniaxial} presents the MD snapshots for layer-selective uniaxial tensile strain, with $U_x$ and $U_y$ applied along the $x$ and $y$ directions, respectively. The strain-free texture in Fig.~\ref{fig:small_angle_uniaxial}(a) provides the reference domain dimensions, with base length $b=132.7~\text{\AA}$ and height $h=114.4~\text{\AA}$. The ratio $h/b$ is close to $\sqrt{3}/2$, indicating that the reference domain is approximately an equilateral triangle.
As the $U_x$ strain increases, the triangular domains in Fig.~\ref{fig:small_angle_uniaxial}(b--d) become progressively narrower and taller. At $\epsilon=1.5\%$, the $h/b$ ratio increases to 1.85.
In contrast, the $U_y$ strain primarily flattens the triangular domains, as shown in Fig.~\ref{fig:small_angle_uniaxial}(e--g). For the same strain magnitude, $U_x$ and $U_y$ produce distinct domain deformations. This difference arises because the two loading directions break the original $C_3$ rotational symmetry in different ways: the $U_x$ loading direction forms an angle with the triangle base, whereas the $U_y$ direction is perpendicular to the base. Figure~\ref{fig:small_angle_uniaxial}(h) compares the $h/b$ values predicted by the registry-mapping model with those obtained from MD simulations. The good agreement indicates that the evolution of the polar topology largely follows the geometric evolution of the stacking registry.

\begin{figure}[t]
\includegraphics[width=1\textwidth]{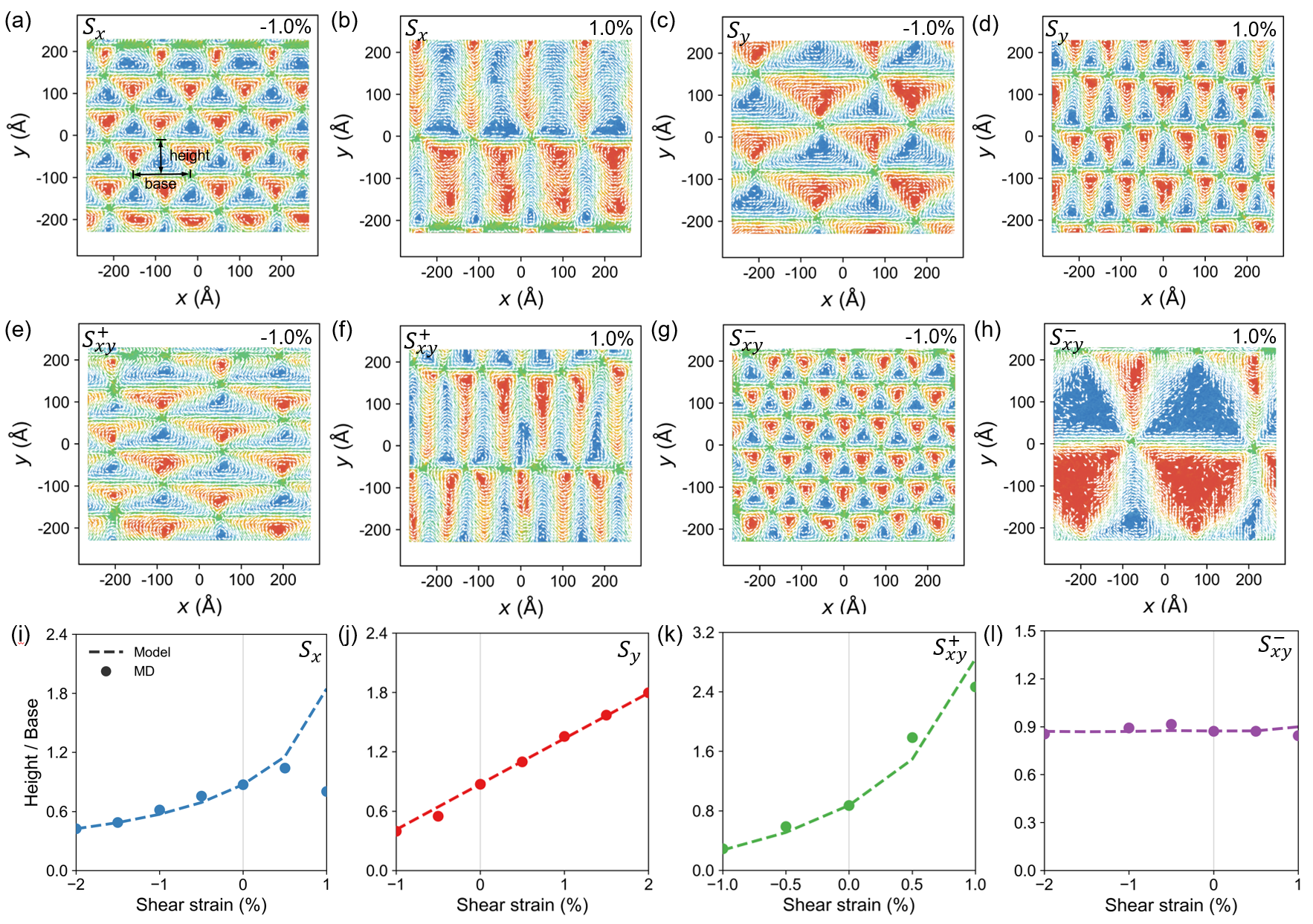}
\caption{\textbf{Layer-selective deformation of triangular polar domains in small-angle twisted bilayer $h$-BN.}
Polarization textures of the $4\times4$ moir\'e supercell with $\theta=1.08^{\circ}$ under layer-selective deformations with amplitudes $\gamma=\pm1.0\%$. The strain-free reference texture is shown in Fig.~\ref{fig:small_angle_uniaxial}(a).
(\textbf{a,b}) $S_x$ deformation with $\gamma=-1.0\%$ and $+1.0\%$, respectively.
(\textbf{c,d}) $S_y$ deformation with $\gamma=-1.0\%$ and $+1.0\%$, respectively.
(\textbf{e,f}) $S_{xy}^{+}$ deformation with $\gamma=-1.0\%$ and $+1.0\%$, respectively.
(\textbf{g,h}) $S_{xy}^{-}$ deformation with $\gamma=-1.0\%$ and $+1.0\%$, respectively.
The background colors denote $P_z$ using the same scale as in Fig.~\ref{fig:polar_texture}(f), and the white arrows
represent the in-plane polarization $\mathbf{P}_{xy}$.
(\textbf{i--l}) Triangular-domain aspect ratio $h/b$ as a function of shear amplitude for $S_x$, $S_y$, $S_{xy}^{+}$, and $S_{xy}^{-}$, respectively.
Solid circles denote MD results extracted from the AA-point geometry, while dashed lines show predictions
of the registry-mapping model.}
\label{fig:small_angle_shear}
\end{figure}

Figure~\ref{fig:small_angle_shear} shows that layer-selective shear produces a more extensive reconstruction of the moir\'e pattern. Compared with the zero-strain pattern in Fig.~\ref{fig:small_angle_uniaxial}(a), $S_x$ deformation changes the number of domain rows along $y$ while largely preserving the domain periodicity along $x$. For $\gamma=-1.0\%$, Fig.~\ref{fig:small_angle_shear}(a) shows that the increased number of rows corresponds to a reduced domain repeat distance along $y$ and laterally elongated domains. For $\gamma=+1.0\%$, Fig.~\ref{fig:small_angle_shear}(b) shows a substantial change in the domain morphology: the triangular domains become closer to rectangular domains.
Figure~\ref{fig:small_angle_shear}(i) compares the $h/b$ values predicted by the registry-mapping model with those obtained from MD simulations. The model agrees well with the MD results for shear strains below $0.5\%$, whereas the deviation at $1.0\%$ coincides with a pronounced change in domain shape from triangular to rectangular.
For $S_y$, the number of rows along $y$ remains nearly unchanged, whereas the number of columns along $x$ decreases for $\gamma=-1.0\%$ [Fig.~\ref{fig:small_angle_shear}(c)] and increases for $\gamma=+1.0\%$ [Fig.~\ref{fig:small_angle_shear}(d)]. Thus, $S_y$ primarily controls the domain repeat distance along $x$. In this case, the registry-mapping model reproduces the nearly linear dependence of $h/b$ on the shear-strain magnitude observed in the MD simulations [Fig.~\ref{fig:small_angle_shear}(j)].

The coupled shear modes modify the two directions simultaneously. Figures~\ref{fig:small_angle_shear}(e) and \ref{fig:small_angle_shear}(f) show that, for $S_{xy}^{+}$, the repeat distances along $x$ and $y$ change in opposite senses when the sign of $\gamma$ is reversed, switching the preferred elongation direction of the triangular domains. For $\gamma=-1.0\%$, the triangular base becomes larger, indicating reduced periodicity along $x$ and increased periodicity along $y$. For $\gamma=+1.0\%$, the opposite trend is observed. The registry-mapping model remains robust for this coupled shear mode and reproduces the corresponding variation of $h/b$ obtained from the MD simulations [Fig.~\ref{fig:small_angle_shear}(k)]. 

By contrast, the row and column counts change together under the rotation-like $S_{xy}^{-}$ mode. As shown in Fig.~\ref{fig:small_angle_shear}(g), negative $\gamma$ produces a denser triangular-domain network, whereas positive $\gamma$ produces a coarser pattern with larger elongated domains and more distorted domain walls [Fig.~\ref{fig:small_angle_shear}(h)]. Despite these pronounced changes in domain size and network density, the $h/b$ ratio remains nearly invariant [Fig.~\ref{fig:small_angle_shear}(l)], indicating that the overall domain aspect ratio is largely preserved. These results show that layer-selective shear can modify both the shape of individual triangular domains and the spacing and arrangement of the domain-wall network.

\section*{Discussion}

Experiments on marginally twisted $h$-BN have revealed charge-polarized triangular moir\'e domains~\cite{woods21p347}. Subsequent studies have reported a variety of distorted moir\'e domain patterns, including elongated triangular domains, stretched domain-wall networks, and stripe-like textures, arising from effects such as flexoelectricity, interlayer bubbles, local tip stress, and mechanically driven interlayer sliding~\cite{Wan24p2410563,Guan25p986}. The diverse domain patterns obtained in our simulations are qualitatively consistent with many of these experimentally observed morphologies, although quantitative agreement is not expected because the loading geometries and boundary conditions may differ. This agreement suggests that layer-selective strain may be realized experimentally, possibly unintentionally, during sample fabrication, transfer, or mechanical manipulation. 

In summary, our large-scale MD simulations reveal that, in twisted bilayer $h$-BN, decreasing the twist angle drives a reconstruction from compact vortex-like textures to extended triangular AB/BA domains. This crossover occurs because lattice relaxation enlarges the low-energy stacking regions and confines polarization reversal to the domain-wall network. The increasing saturated-domain area fraction quantitatively captures the expansion of regions with weak in-plane polarization and nearly saturated out-of-plane polarization, while the spatial distribution of $|\mathbf{P}_{xy}|$ reveals their reorganization into triangular domains in the small-angle regime. Across this morphological crossover, the mean absolute in-plane winding remains close to unity up to $1000~\mathrm{K}$.

We find that the mechanical response is governed by the relative deformation of the two layers through the resulting change in the interlayer registry field. We show that layer-selective deformation, which has remained largely unexplored in previous studies, provides an effective knob for modifying moir\'e polar patterns. By directly altering the relative stacking field, layer-selective deformation reshapes both the large-angle vortex lattice and the small-angle triangular-domain network. The registry-mapping model provides a unified geometric description of these responses across different polar morphologies, capturing the deformation-induced changes in the directional vortex radii as well as in the base and height of triangular domains. These results demonstrate relative layer deformation as a powerful geometric route to mechanically reshape moir\'e polarization textures.

\section*{Methods}

\subsection*{First-principles calculations and deep potential model}

A deep potential (DP) model \cite{Wang18p178} was employed to enable large-scale molecular dynamics simulations. The reference energies and atomic forces were obtained from density functional theory (DFT) calculations performed using the Vienna \textit{Ab initio} Simulation Package (\texttt{VASP}) \cite{Kresse96p11169,Kresse96p15}. 
The projector augmented-wave method \cite{Blochl94p17953} and the Perdew--Burke--Ernzerhof functional \cite{Perdew96p3865} were used to describe the electron--ion and exchange--correlation interactions, respectively. Long-range dispersion was included using the Grimme DFT-D2 correction \cite{Grimme06p1787}. A plane-wave cutoff energy of 600~eV and a $\Gamma$-centered $k$-point mesh with a reciprocal-space spacing of 0.25~\AA$^{-1}$ were employed. The electronic self-consistency criterion was $10^{-7}$~eV. A vacuum spacing of 20~\AA\ was introduced along the out-of-plane direction, and a dipole correction was applied along this direction \cite{Neugebauer92p16067}. Symmetry was disabled in all calculations, and the reference energies and forces were obtained from static single-point calculations.

The smooth DeepPot-SE scheme~\cite{Zhang18p4436,Zhang18p143001} was used to represent the potential-energy surface. The DP model employed in this work was developed and validated in a companion study available as a public preprint~\cite{fan25p2510.13568}; the same parameterization was used here without further retraining. Briefly, the model was trained against DFT energies and atomic forces for bilayer $h$-BN configurations covering AB, BA, and AA stackings, various in-plane relative displacements, interlayer separations, and applied strains. The training database contained 443 configurations after cleaning and was subsequently expanded to 1,513 configurations using the DP-GEN concurrent-learning protocol~\cite{Zhang20p107206,zhang19p023804}. A compact $64\times64$ neural-network architecture with model compression was adopted for efficient large-scale simulations~\cite{Zeng23p054801}. On the DFT benchmark, the model gives mean absolute errors of 0.507~meV/atom for energies and 0.024~eV/\AA\ for atomic forces, and reproduces the DFT phonon spectra, strain energetics, and interlayer sliding barriers.

\subsection*{Deep-potential molecular dynamics simulations}

Commensurate moir\'e superlattices were constructed for twist indices from $i=5$ to $i=30$. The upper and lower layers were rotated by $+\theta/2$ and $-\theta/2$, respectively, so that their relative twist is $\theta$. The commensurate angle satisfies
\begin{equation}
\cos\theta=
\frac{3i^2+3i+0.5}{3i^2+3i+1}.
\end{equation}
The deformation simulations used an $i=5$ structure ($\theta=6.01^{\circ}$; 364 atoms per moir\'e unit cell)
and an $i=30$ structure ($\theta=1.08^{\circ}$; 11{,}164 atoms per moir\'e unit cell). A $6\times6$ moir\'e supercell was used for the $6.01^{\circ}$ structure, and a $4\times4$ moir\'e supercell was used for the $1.08^{\circ}$ structure.
All molecular dynamics simulations were performed using \texttt{LAMMPS}~\cite{Plimpton95p1} with a time step of 1~fs. The structures were first equilibrated at $300~\mathrm{K}$ for 500~ps in the $NPT$ ensemble using a Nos\'e--Hoover thermostat and barostat~\cite{Hoover85p1695}. Only the in-plane simulation-cell vectors were allowed to relax, at zero in-plane pressure. The resulting equilibrium structures were subsequently evolved in the $NVT$ ensemble for 300~ps. Unless otherwise specified, all $NVT$ simulations were performed at 300~K. All deformation simulations were subsequently performed in the $NVT$ ensemble with the simulation cell fixed.

Layer-selective deformation was imposed on the upper layer about its geometric center, while the lower layer and simulation cell were kept unchanged. Writing $\tilde{\mathbf r}=(x-x_c,y-y_c)^{\mathrm T}$, the centered upper-layer coordinates were transformed as $\tilde{\mathbf r}'=\mathbf F\tilde{\mathbf r}$, with
\begin{equation}
\mathbf F_{S_x}=
\begin{pmatrix}
1 & \gamma\\
0 & 1
\end{pmatrix},
\quad
\mathbf F_{S_y}=
\begin{pmatrix}
1 & 0\\
\gamma & 1
\end{pmatrix},
\quad
\mathbf F_{S_{xy}^{\pm}}=
\begin{pmatrix}
1 & \gamma\\
\pm\gamma & 1
\end{pmatrix},
\end{equation}
and
\begin{equation}
\mathbf F_{U_x}=
\begin{pmatrix}
1+\epsilon & 0\\
0 & 1
\end{pmatrix},
\quad
\mathbf F_{U_y}=
\begin{pmatrix}
1 & 0\\
0 & 1+\epsilon\\
\end{pmatrix}.
\end{equation}
Here, $(x_c,y_c)$ is the geometric center of the upper layer, $\gamma$ is the signed shear amplitude, and $\epsilon$ is the engineering strain. 
These deformations disrupt the original periodic matching across the $y$ boundaries for $S_x$ and $U_y$, across the $x$ boundaries for $S_y$ and $U_x$, and across both boundaries for $S_{xy}^{\pm}$. Accordingly, the bilayer was treated as finite along the direction in which periodic matching was lost: finite in the $y$ direction for $S_x$ and $U_y$, finite in the $x$ direction for $S_y$ and $U_x$, and finite in both directions for $S_{xy}^{\pm}$. For these configurations, atoms within nominally 7~\AA-wide strips at the corresponding boundaries were held fixed during the $NVT$ simulations to stabilize the imposed finite deformation.

To assess thermal stability, simulations were performed at $5~\mathrm{K}$ and at temperatures from $100$ to $1000~\mathrm{K}$ in $100~\mathrm{K}$ increments. At each temperature, the system was evolved for 300~ps, and topological observables were averaged over configurations sampled during the final $100~\mathrm{ps}$, while representative polarization textures were obtained from configurations sampled during the same interval. The local polarization was evaluated from atomic displacements weighted by registry-dependent Born effective charges and normalized by the local unit-cell area, following the procedure established in Ref.~\cite{ke25p046201}. The saturated-domain area fraction was defined as the fraction of the moir\'e unit cell satisfying $|\mathbf{P}_{xy}|<0.40\times10^{-3}$~$e$/\AA~and $|P_z|>1.20$~$e$/\AA. To characterize triangular-domain geometry, AA centers were identified from the local interlayer registry. The domain base $b$ was defined by one continuously tracked AA--AA neighbor edge, and the corresponding height $h$ was evaluated as the perpendicular distance from the third AA point to this edge. The same AA-neighbor branch was used for the strain-free and deformed structures.

\vspace{1cm}

{\bf{Data Availability}} The data generated and analyzed in this study are available from the corresponding author upon reasonable request.

{\bf{Acknowledgments}} D.F., C.K. and S.L. acknowledge the support from National Natural Science Foundation of China (Grants No.~12304128 and~12674120) and Zhejiang Provincial Natural Science Foundation of China (Grants No.~LR25A040004 and DG26A04000). The computational resource is provided by Westlake HPC Center.

{\bf{Author Contributions}} S.L. and C.K. conceived the idea and designed the project. D.F. carried out the calculations and wrote a major part of the paper. S.L. and C.K. revised the paper. All authors reviewed and edited the manuscript.

{\bf{Competing Interests}} The authors declare no competing financial or non-financial interests.

\bibliography{SL}

@misc{fan25p2510.13568,
      title={Rippled Moire Superlattices for Decoupled Ferroelectric Bits}, 
      author={Di Fan and Changming Ke and Shi Liu},
      year={2025},
      eprint={2510.13568},
      archivePrefix={arXiv},
      primaryClass={physics.comp-ph},
      url={https://arxiv.org/abs/2510.13568}, 
}

@article{Hong26p1022,
  author  = {Hong, Zijian and Shao, Yu-Tsun and Tian, He and Roychowdhury, Krishanu and Das, Sujit},
  title   = {Topological textures in polar materials},
  journal = {Nature Physics},
  volume  = {22},
  pages   = {1022--1030},
  year    = {2026},
  doi     = {10.1038/s41567-026-03344-x}
}

@article{Bistritzer11p12233,
  author  = {Bistritzer, R. and MacDonald, A. H.},
  title   = {Moir{\'e} bands in twisted double-layer graphene},
  journal = {Proceedings of the National Academy of Sciences},
  volume  = {108},
  pages   = {12233--12237},
  year    = {2011},
  doi     = {10.1073/pnas.1108174108}
}

@article{Andrei20p1265,
  author  = {Andrei, E. Y. and MacDonald, A. H.},
  title   = {Graphene bilayers with a twist},
  journal = {Nature Materials},
  volume  = {19},
  pages   = {1265--1275},
  year    = {2020},
  doi     = {10.1038/s41563-020-00840-0}
}

@article{Seyler19p66,
  author  = {Seyler, K. L. and others},
  title   = {Signatures of moir{\'e}-trapped valley excitons in MoSe$_2$/WSe$_2$ heterobilayers},
  journal = {Nature},
  volume  = {567},
  pages   = {66--70},
  year    = {2019},
  doi     = {10.1038/s41586-019-0957-1}
}

@article{Ni19p4360,
  author  = {Ni, G. X. and others},
  title   = {Soliton superlattices in twisted hexagonal boron nitride},
  journal = {Nature Communications},
  volume  = {10},
  pages   = {4360},
  year    = {2019},
  doi     = {10.1038/s41467-019-12327-x}
}

@article{Kim24p65,
  author  = {Kim, D. S. and others},
  title   = {Electrostatic moir{\'e} potential from twisted hexagonal boron nitride layers},
  journal = {Nature Materials},
  volume  = {23},
  pages   = {65--70},
  year    = {2024},
  doi     = {10.1038/s41563-023-01637-7}
}

@article{Wan24p2410563,
  author  = {Wan, S. and others},
  title   = {Intertwined flexoelectricity and stacking ferroelectricity in marginally twisted hBN moir{\'e} superlattice},
  journal = {Advanced Materials},
  volume  = {36},
  pages   = {2410563},
  year    = {2024},
  doi     = {10.1002/adma.202410563}
}

@article{Bennett23p1629,
  author  = {Bennett, D. and Chaudhary, G. and Slager,R.-J. and others},
  title   = {Polar meron-antimeron networks in strained and twisted bilayers},
  journal = {Nature Communications},
  volume  = {14},
  pages   = {1629},
  year    = {2023},
  doi     = {10.1038/s41467-023-37337-8}
}

@article{Pan25p3026,
  author  = {Pan, Er and Li, Zefen and Yang, Fan and Niu, Kangdi and Bian, Renji and Liu, Qing and Chen, Jiangang and Dong, Biao and Wang, Ruixue and Zhou, Tingchuan and Zhou, Aijun and Luo, Xiao and Chu, Junhao and Lin, Junhao and Li, Wenwu and Liu, Fucai},
  title   = {Observation and manipulation of two-dimensional topological polar texture confined in moir{\'e} interface},
  journal = {Nature Communications},
  volume  = {16},
  pages   = {3026},
  year    = {2025},
  doi     = {10.1038/s41467-025-58105-w}
}

@article{Ma24p224105,
  author  = {Ma, W. and Yuan, S. and Liu, Y. and Wang, B.},
  title   = {Polar vortex generation and control in a ferroelectric nanodisk},
  journal = {Physical Review B},
  volume  = {110},
  pages   = {224105},
  year    = {2024},
  doi     = {10.1103/PhysRevB.110.224105}
}

@article{Guan25p986,
  author = {Guan, Zhao and Wei, Lu-qi and Fan, Wen-cheng and Sun, Yi-chen and Cao, Wei and Tian, Ming and Wan, Neng and Tong, Wen-yi and Chen, Bin-bin and Xiang, Ping-hua and Duan, Chun-gang and Zhong, Ni},
  title = {Mechanical force-induced interlayer sliding in interfacial ferroelectrics},
  journal = {Nature Communications},
  volume = {16},
  pages = {986},
  year = {2025},
  doi = {10.1038/s41467-025-56073-9}
}

@article{moore21p5741,
  title={Nanoscale lattice dynamics in hexagonal boron nitride moir{\'e} superlattices},
  author={Moore, SL and Ciccarino, CJ and Halbertal, Dorri and McGilly, LJ and Finney, NR and Yao, Kaiyuan and Shao, Yinming and Ni, Guangxin and Sternbach, Aaron and Telford, EJ and others},
  journal={Nat. Commun.},
  volume={12},
  number={1},
  pages={5741},
  year={2021},
  doi = {10.1038/s41467-021-26072-7},
  url = {https://doi.org/10.1038/s41467-021-26072-7},
  publisher={Nature Publishing Group UK London}
}

@article{Bennett22p235445,
  title = {Theory of polar domains in moir\'e heterostructures},
  author = {Bennett, Daniel},
  journal = {Phys. Rev. B},
  volume = {105},
  issue = {23},
  pages = {235445},
  numpages = {14},
  year = {2022},
  month = {Jun},
  publisher = {American Physical Society},
  doi = {10.1103/PhysRevB.105.235445},
  url = {https://link.aps.org/doi/10.1103/PhysRevB.105.235445}
}

@article{li25p5451,
  title={Unusual topological polar texture in moir{\'e} ferroelectrics},
  author={Li, Yuhao and Wei, Yuanhao and Guo, Ruiping and Wang, Yifei and Zhang, Hanhao and Taniguchi, Takashi and Watanabe, Kenji and Shi, Yan and Shi, Yi and Wang, Chong and others},
  journal={Nat. Commun.},
  volume={16},
  number={1},
  pages={5451},
  year={2025},
  doi = {10.1038/s41467-025-60647-y},
  url = {https://doi.org/10.1038/s41467-025-60647-y},
  publisher={Nature Publishing Group UK London}
}

@article{Bennett22p7,
  title = {On electrically tunable stacking domains and ferroelectricity in moiré superlattices},
  author={Bennett, Daniel and Remez, Benjamin},
  journal={npj 2D Mater. Appl.},
  volume={6},
  number={1},
  pages={7},
  year={2022},
  publisher={Nature Publishing Group UK London},
  doi = {10.1038/s41699-021-00281-6},
}

@article{ke25p046201,
  title = {Superlubric Motion of Wavelike Domain Walls in Sliding Ferroelectrics},
  author = {Ke, Changming and Liu, Fucai and Liu, Shi},
  journal = {Phys. Rev. Lett.},
  volume = {135},
  issue = {4},
  pages = {046201},
  numpages = {7},
  year = {2025},
  month = {Jul},
  publisher = {American Physical Society},
  doi = {10.1103/jhlq-2dd7},
  url = {https://link.aps.org/doi/10.1103/jhlq-2dd7}
}

@article{woods21p347,
  title={Charge-polarized interfacial superlattices in marginally twisted hexagonal boron nitride},
  author={Woods, CR and Ares, Pablo and Nevison-Andrews, Harriet and Holwill, MJ and Fabregas, Rene and Guinea, Francisco and Geim, AK and Novoselov, KS and Walet, NR and Fumagalli, Laura},
  journal={Nat. Commun.},
  volume={12},
  number={1},
  pages={347},
  year={2021},
  doi= {10.1038/s41467-020-20667-2},
  url ={https://doi.org/10.1038/s41467-020-20667-2},
  publisher={Nature Publishing Group UK London}
}

@article{Shao2023p1355,
  title={Emergent chirality in a polar meron to skyrmion phase transition},
  author={Shao, Yu-Tsun and Das, Sujit and Hong, Zijian and Xu, Ruijuan and Chandrika, Swathi and G{\'o}mez-Ortiz, Fernando and Garc{\'\i}a-Fern{\'a}ndez, Pablo and Chen, Long-Qing and Hwang, Harold Y and Junquera, Javier and others},
  journal={Nat. Commun.},
  url = {https://doi.org/10.1038/s41467-023-36950-x},
  DOI = {10.1038/s41467-023-36950-x},
  volume={14},
  number={1},
  pages={1355},
  year={2023},
  publisher={Nature Publishing Group UK London}
}

@article{Junquera23p025001,
  title={Topological phases in polar oxide nanostructures},
  author={Junquera, Javier and Nahas, Yousra and Prokhorenko, Sergei and Bellaiche, Laurent and {\'I}{\~n}iguez, Jorge and Schlom, Darrell G and Chen, Long-Qing and Salahuddin, Sayeef and Muller, David A and Martin, Lane W and others},
  journal={Rev. Mod. Phys.},
  volume={95},
  number={2},
  pages={025001},
  year={2023},
  publisher={APS}
}

@article{Blochl94p17953,
  title = {Projector Augmented-Wave Method},
  author = {Bl{\"o}chl, P. E.},
  year = {1994},
  month = dec,
  journal = {Phys. Rev. B},
  volume = {50},
  number = {24},
  pages = {17953--17979},
  publisher = {American Physical Society},
  doi = {10.1103/PhysRevB.50.17953}
}

@article{Zeng23p054801,
	author = {Zeng, Jinzhe and Zhang, Duo and Lu, Denghui and Mo, Pinghui and Li, Zeyu and Chen, Yixiao and Rynik, Mari{\'a}n and Huang, Li'ang and Li, Ziyao and Shi, Shaochen and Wang, Yingze and Ye, Haotian and Tuo, Ping and Yang, Jiabin and Ding, Ye and Li, Yifan and Tisi, Davide and Zeng, Qiyu and Bao, Han and Xia, Yu and Huang, Jiameng and Muraoka, Koki and Wang, Yibo and Chang, Junhan and Yuan, Fengbo and Bore, Sigbj{\o}rn L{\o}land and Cai, Chun and Lin, Yinnian and Wang, Bo and Xu, Jiayan and Zhu, Jia-Xin and Luo, Chenxing and Zhang, Yuzhi and Goodall, Rhys E. A. and Liang, Wenshuo and Singh, Anurag Kumar and Yao, Sikai and Zhang, Jingchao and Wentzcovitch, Renata and Han, Jiequn and Liu, Jie and Jia, Weile and York, Darrin M. and E, Weinan and Car, Roberto and Zhang, Linfeng and Wang, Han},
	issn = {0021-9606},
	journal = {J. Chem. Phys.},
	month = {08},
	number = {5},
	pages = {054801},
	title = {{DeePMD-kit v2: A software package for deep potential models}},
	volume = {159},
	year = {2023},
        doi = {10.1063/5.0155600}
}

@article{Zhang18p4436,
	author = {Zhang, Linfeng and Han, Jiequn and Wang, Han and Saidi, Wissam A. and Car, Roberto and E, Weinan},
	journal = {Advances in Neural Information Processing Systems},
	month = 12,
	pages = {4436-4446},
    url = {https://dl.acm.org/doi/abs/10.5555/3327345.3327356},
	title = {End-to-end symmetry preserving inter-atomic potential energy model for finite and extended systems},
	volume = {2018}}

@article{Geim13p419,
	author = {Geim, A. K. and Grigorieva, I. V.},
	journal = {Nature},
	number = {7459},
	pages = {419-425},
    url = {https://doi.org/10.1038/nature12385},
    DOI = {10.1038/nature12385},
	title = {Van der Waals heterostructures},
	volume = {499},
	year = {2013}}

@article{Hoover85p1695,
	author = {W. G. Hoover},
	journal = {Phys. Rev. A},
	pages = {1695--7},
	title = {Canonical dynamics: Equilibrium phase-space distributions},
	volume = {31},
        doi = {10.1103/PhysRevA.31.1695},
	year = {1985}}

@article{Das21p194,
	author = {Das, Sujit and Hong, Zijian and Stoica, VA and Gon{\c{c}}alves, MAP and Shao, Yu-Tsun and Parsonnet, Eric and Marksz, Eric J and Saremi, Sahar and McCarter, MR and Reynoso, A and others},
	journal = {Nat. Mater.},
    url = {https://doi.org/10.1038/s41563-020-00818-y},
    DOI = {10.1038/s41563-020-00818-y},
	number = {2},
	pages = {194--201},
	publisher = {Nature Publishing Group},
	title = {Local negative permittivity and topological phase transition in polar skyrmions},
	volume = {20},
	year = {2021}}

@article{Cao18p80,
  doi = {10.1038/nature26154},
  url = {https://doi.org/10.1038/nature26154},
  year = {2018},
  month = mar,
  publisher = {Springer Science and Business Media {LLC}},
  volume = {556},
  number = {7699},
  pages = {80--84},
  author = {Yuan Cao and Valla Fatemi and Ahmet Demir and Shiang Fang and Spencer L. Tomarken and Jason Y. Luo and Javier D. Sanchez-Yamagishi and Kenji Watanabe and Takashi Taniguchi and Efthimios Kaxiras and Ray C. Ashoori and Pablo Jarillo-Herrero},
  title = {Correlated insulator behaviour at half-filling in magic-angle graphene superlattices},
  journal = {Nature}
}

@article{Zheng20p71,
  doi = {10.1038/s41586-020-2970-9},
  url = {https://doi.org/10.1038/s41586-020-2970-9},
  year = {2020},
  month = nov,
  publisher = {Springer Science and Business Media {LLC}},
  volume = {588},
  number = {7836},
  pages = {71--76},
  author = {Zhiren Zheng and Qiong Ma and Zhen Bi and Sergio de la Barrera and Ming-Hao Liu and Nannan Mao and Yang Zhang and Natasha Kiper and Kenji Watanabe and Takashi Taniguchi and Jing Kong and William A. Tisdale and Ray Ashoori and Nuh Gedik and Liang Fu and Su-Yang Xu and Pablo Jarillo-Herrero},
  title = {Unconventional ferroelectricity in moir{\'{e}} heterostructures},
  journal = {Nature}
}

@article{Tran19p71,
  title = {Evidence for moiré excitons in van der Waals heterostructures},
  volume = {567},
  ISSN = {1476-4687},
  url = {http://dx.doi.org/10.1038/s41586-019-0975-z},
  DOI = {10.1038/s41586-019-0975-z},
  number = {7746},
  journal = {Nature},
  publisher = {Springer Science and Business Media LLC},
  author = {Tran,  Kha and Moody,  Galan and Wu,  Fengcheng and Lu,  Xiaobo and Choi,  Junho and Kim,  Kyounghwan and Rai,  Amritesh and Sanchez,  Daniel A. and Quan,  Jiamin and Singh,  Akshay and Embley,  Jacob and Zepeda,  André and Campbell,  Marshall and Autry,  Travis and Taniguchi,  Takashi and Watanabe,  Kenji and Lu,  Nanshu and Banerjee,  Sanjay K. and Silverman,  Kevin L. and Kim,  Suenne and Tutuc,  Emanuel and Yang,  Li and MacDonald,  Allan H. and Li,  Xiaoqin},
  year = {2019},
  month = feb,
  pages = {71-75}
}

@Article{Kresse96p15,
  author    = {G. Kresse and J. Furthm\"{u}ller},
  journal   = {Comput. Mater. Sci.},
  title     = {Efficiency of ab-initio total energy calculations for metals and semiconductors using a plane-wave basis set},
  year      = {1996},
  month     = {jul},
  number    = {1},
  pages     = {15--50},
  volume    = {6},
  doi       = {10.1016/0927-0256(96)00008-0},
  publisher = {Elsevier {BV}}
}

@article{Zhang19p023804,
	author = {Zhang, Linfeng and Lin, De-Ye and Wang, Han and Car, Roberto and E, Weinan},
	journal = {Phys. Rev. Mater.},
	pages = {023804},
	title = {{Active learning of uniformly accurate interatomic potentials for materials simulation}},
	volume = {3},
        DOI = {10.1103/PhysRevMaterials.3.023804},
	year = {2019}}

@Article{Grimme06p1787,
  author    = {Stefan Grimme},
  journal   = {J. Comput. Chem.},
  title     = {Semiempirical {GGA}-type density functional constructed with a long-range dispersion correction},
  year      = {2006},
  number    = {15},
  pages     = {1787--1799},
  volume    = {27},
  doi       = {10.1002/jcc.20495},
  publisher = {Wiley},
}

@Article{Perdew96p3865,
  author    = {Perdew, John P. and Burke, Kieron and Ernzerhof, Matthias},
  journal   = {Phys. Rev. Lett.},
  title     = {Generalized Gradient Approximation Made Simple},
  year      = {1996},
  month     = {Oct},
  pages     = {3865--3868},
  volume    = {77},
  doi       = {10.1103/PhysRevLett.77.3865},
  issue     = {18},
  numpages  = {0},
  publisher = {American Physical Society},
}

@article{Neugebauer92p16067,
  title = {Adsorbate-substrate and adsorbate-adsorbate interactions of \text{Na} and \text{K} adlayers on \text{Al}(111)},
  author = {Neugebauer, J\"org and Scheffler, Matthias},
  journal = {Phys. Rev. B},
  volume = {46},
  issue = {24},
  pages = {16067-16080},
  year = {1992},
  month = {Dec},
  publisher = {American Physical Society},
  doi = {10.1103/PhysRevB.46.16067},
}

@article{Kresse96p11169,
	author = {Kresse, G and Furthm{\"u}ller, J},
	journal = {Phys. Rev. B},
	number = {16},
	pages = {11169-11186},
	title = {Efficient iterative schemes for \textit{ab initio} total-energy calculations using a plane-wave basis set},
	volume = {54},
	year = {1996},
    doi = {10.1103/PhysRevB.54.11169}}

@Article{Li17p6382,
  author    = {Lei Li and Menghao Wu},
  journal   = {{ACS} Nano},
  title     = {Binary Compound Bilayer and Multilayer with Vertical Polarizations: Two-Dimensional Ferroelectrics, Multiferroics, and Nanogenerators},
  year      = {2017},
  month     = {jun},
  number    = {6},
  pages     = {6382--6388},
  volume    = {11},
  doi       = {10.1021/acsnano.7b02756},
url = {https://doi.org/10.1021/acsnano.7b02756},
  publisher = {American Chemical Society ({ACS})},
}

@Article{Wang18p178,
  author    = {Han Wang and Linfeng Zhang and Jiequn Han and Weinan E},
  journal   = {Comput. Phys. Commun.},
  title     = {{DeePMD}-kit: A deep learning package for many-body potential energy representation and molecular dynamics},
  year      = {2018},
  month     = jul,
  pages     = {178--184},
  volume    = {228},
  doi       = {10.1016/j.cpc.2018.03.016},
  publisher = {Elsevier {BV}},
}

@Article{Cao18p43,
  author    = {Yuan Cao and Valla Fatemi and Shiang Fang and Kenji Watanabe and Takashi Taniguchi and Efthimios Kaxiras and Pablo Jarillo-Herrero},
  journal   = {Nature},
  title     = {Unconventional superconductivity in magic-angle graphene superlattices},
  year      = {2018},
  month     = mar,
  number    = {7699},
  pages     = {43--50},
  volume    = {556},
  doi       = {10.1038/nature26160},
  publisher = {Springer Science and Business Media {LLC}},
}

@Article{ViznerStern21p1462,
  author    = {M. Vizner Stern and Y. Waschitz and W. Cao and I. Nevo and K. Watanabe and T. Taniguchi and E. Sela and M. Urbakh and O. Hod and M. Ben Shalom},
  journal   = {Science},
  title     = {Interfacial ferroelectricity by van der Waals sliding},
  year      = {2021},
  month     = jun,
  number    = {6549},
  pages     = {1462--1466},
  volume    = {372},
  doi       = {10.1126/science.abe8177},
  publisher = {American Association for the Advancement of Science ({AAAS})},
}

@Article{Zhang18p143001,
  author    = {L. Zhang and J. Han and H. Wang and R. Car and W. E},
  journal   = {Phys. Rev. Lett.},
  title     = {Deep Potential Molecular Dynamics: A Scalable Model with the Accuracy of Quantum Mechanics},
  year      = {2018},
  month     = {apr},
  number    = {14},
  pages     = {143001},
  volume    = {120},
  doi       = {10.1103/PhysRevLett.120.143001},
  publisher = {American Physical Society ({APS})},
}

@Article{Plimpton95p1,
  author    = {Plimpton, Steve},
  journal   = {J. Comput. Phys.},
  title     = {Fast Parallel Algorithms for Short-Range Molecular Synamics},
  year      = {1995},
  number    = {1},
  pages     = {1--19},
  volume    = {117},
  doi       = {10.1006/jcph.1995.1039},
  publisher = {Elsevier},
}

@article{Wang22p367,
	title = {Interfacial ferroelectricity in rhombohedral-stacked bilayer transition metal dichalcogenides},
        doi = {10.1038/s41565-021-01059-z},
	volume = {17},
	journal = {Nat. Nanotechnol.},
	author = {Wang, XR and Yasuda, K and Zhang, Y and Liu, S and Watanabe, K and Taniguchi, T and Hone, J and Fu, L and Jarillo-Herrero, P},
	year = {2022},
	pages = {367--372},
}

@article{Weston22p390,
	title = {Interfacial ferroelectricity in marginally twisted {2D} semiconductors},
	volume = {17},
	url = {https://www.nature.com/articles/s41565-022-01072-w},
	number = {4},
	journal = {Nat. Nanotechnol.},
	author = {Weston, Astrid and Castanon, Eli G. and Enaldiev, Vladimir and Ferreira, Fábio and Bhattacharjee, Shubhadeep and Xu, Shuigang and Corte-León, Héctor and Wu, Zefei and Clark, Nicholas and Summerfield, Alex and Hashimoto, Teruo and Gao, Yunze and Wang, Wendong and Hamer, Matthew and Read, Harriet and Fumagalli, Laura and Kretinin, Andrey V. and Haigh, Sarah J. and Kazakova, Olga and Geim, A. K. and Fal’ko, Vladimir I. and Gorbachev, Roman},
	year = {2022},
	pages = {390--395}
}

@Article{Yasuda21p1458,
  author    = {Yasuda, Kenji and Wang, Xirui and Watanabe, Kenji and Taniguchi, Takashi and Jarillo-Herrero, Pablo},
  journal   = {Science},
  title     = {Stacking-engineered ferroelectricity in bilayer boron nitride},
  year      = {2021},
  issn      = {1095-9203},
  month     = jun,
  number    = {6549},
  pages     = {1458–1462},
  volume    = {372},
  doi       = {10.1126/science.abd3230},
  publisher = {American Association for the Advancement of Science (AAAS)},
}

@Article{Zhang20p107206,
  author    = {Yuzhi Zhang and Haidi Wang and Weijie Chen and Jinzhe Zeng and Linfeng Zhang and Han Wang and Weinan E},
  journal   = {Comput. Phys. Commun.},
  title     = {{DP}-{GEN}: A concurrent learning platform for the generation of reliable deep learning based potential energy models},
  year      = {2020},
  month     = aug,
  pages     = {107206},
  volume    = {253},
  doi       = {10.1016/j.cpc.2020.107206},
  publisher = {Elsevier {BV}},
}

\clearpage

\clearpage

\clearpage

\clearpage

\clearpage

\clearpage

\clearpage

\end{document}